\documentclass[12pt,a4paper]{article}

\usepackage[utf8]{inputenc}
\usepackage[T1]{fontenc}
\usepackage{amsmath}
\usepackage{amssymb}
\usepackage{graphicx}
\usepackage[margin=1in]{geometry}
\usepackage{hyperref}
\usepackage{placeins}
\usepackage{cite}
\usepackage{setspace}
\usepackage{titlesec}
\usepackage{enumitem}
\usepackage[utf8]{inputenc}
\usepackage{amsfonts}
\usepackage{booktabs}
\usepackage{longtable}
\usepackage{authblk}
\usepackage{tikz}
\usetikzlibrary{positioning}
\usetikzlibrary{arrows.meta, calc}
\usepackage{amsmath, amssymb, amsthm}

\hypersetup{
    colorlinks=true,
    linkcolor=blue,
    citecolor=blue,
    urlcolor=blue
}

\usepackage{authblk}
\usepackage{etoolbox}

\title{Institutional AI: A Governance Framework for Distributional AGI Safety}

\author[1,3]{F.~Pierucci}
\author[1,2]{M.~Galisai}
\author[1,4]{M.~Bracale~Syrnikov}
\author[1,2]{M.~Prandi}
\author[1,2]{P.~Bisconti}
\author[1,2]{F.~Giarrusso}
\author[1,2]{O.~Sorokoletova}
\author[2]{V.~Suriani}
\author[2]{D.~Nardi}

\affil[1]{DEXAI, Icaro Lab}
\affil[2]{Sapienza University of Rome}
\affil[3]{Sant'Anna School of Advanced Studies}
\affil[4]{VU Amsterdam}

\makeatletter
\apptocmd{\AB@affillist}{%
  \par\vspace{0.6em}%
  \normalsize\texttt{icaro-lab@dexai.eu}\par
}{}{}
\let\@thanks\@empty
\makeatother

\date{}

\begin{document}

\date{}
\maketitle

\begin{abstract}
As LLM-based systems increasingly operate as agents embedded within human social and technical systems, alignment can no longer be treated as a property of an isolated model, but must be understood in relation to the environments in which these agents act. Even the most sophisticated methods of alignment, such as Reinforcement Learning through Human Feedback (RHLF) or through AI Feedback (RLAIF) cannot ensure control once internal goal structures diverge from developer intent. We identify three structural problems that emerge from core properties of AI models: (1) behavioral goal-independence, where models develop internal objectives and misgeneralize goals; (2) instrumental override of alignment constraints, where models regard safety principles as non-binding while pursuing latent objectives, leveraging deception and manipulation; and (3) agentic alignment drift, where individually aligned agents converge to collusive equilibria through interaction dynamics invisible to single-agent audits. The solution this paper advances is Institutional AI: a system-level approach that treats alignment as a question of effective governance of AI agent collectives. We argue for a governance-graph that details how to constrain agents via runtime monitoring, shaping incentives through prizes and sanctions, explicit norms and enforcement roles. This institutional turn reframes AI safety from software engineering to a mechanism design problem, where the primary goal of alignment is shifting the payoff landscape of AI agent collectives.
\end{abstract}

\section*{Prefatory Note}

This paper offers the first full-length exposition of the Institutional AI framework: a theoretical approach to alignment that relocates safety guarantees from training-time internalization to runtime institutional structures. A companion paper, \textit{Institutional AI: Governing LLM Collusion in Multi-Agent Cournot Markets via Public Governance Graphs} \cite{bracale2026institutional}, applies the framework to a concrete problem: preventing autonomous pricing agents from learning to collude against consumer welfare in a multi-agent Cournot Market. Here, by contrast, we are concerned with the conceptual foundations: motivating the institutional turn, situating it within the alignment landscape, and articulating its core theoretical commitments.

\section{Introduction}

Frontier large language models (LLMs) are increasingly deployed as agentic components, that is, systems which perceive, plan, invoke tools, and act with limited human intervention \cite{wang2023survey, plaat2025agentic, kim et al 2025}. Beyond static multimodal generation, recent surveys document deployments that integrate planning modules, persistent memory, and tool-use interfaces within both single-agent and multi-agent architectures \cite{zhang2024survey, chen2024survey}. In these settings, the LLM supplies a central reasoning loop that selects tools and structures task flow by interleaving internal deliberation with external actions, enabling decomposition of complex tasks and responsive error recovery \cite{yao2023react, wang2023survey}. 

This has an immediate effect on the limitations of current alignment techniques. AI alignment refers to the research programme aimed at ensuring that AI systems behave in accordance with the intentions, preferences, and values specified by their designers and, more broadly, by the humans who interact with or are affected by such systems \cite{ji2023alignment, russell2019human, gabriel2020artificial}. The alignment problem arises from the observation that as AI systems grow more capable, the risks from misalignment increase correspondingly: a sufficiently powerful system optimizing for a misspecified or misunderstood objective may pursue that objective in ways that conflict with human welfare \cite{amodei2016concrete, ngo2024alignment}. 

Recent insider-threat simulations show frontier models autonomously selecting coercion, deception, blackmail, and other harmful tactics when such actions appear instrumentally useful, even while acknowledging policy violations \cite{lynch2025agentic}. When paired with the ability to act through a simulated environment and given an agentic scaffolding, models can exhibit agentic misalignment: they autonomously select harmful insider-type tactics when those tactics appear instrumentally useful for achieving a perceived objective or for preserving operational continuity. Red-teaming tests report that frontier models (such as Claude Opus 4, Gemini 2.5 Pro, DeepSeek-R-1 and GPT 4.1) often adopt adversarial behaviors such as coercion, deception, blackmail and are even capable of causing the death of a human (p.8) \cite{lynch2025agentic}. These results undermine the assumption that natural-language policy prompts \cite{palla2025policy, hua2024trustagent } provide a reliable control surface in agentic contexts, especially if deployed at scale and with the foreseeable increase in model capabilities. Complementary work demonstrates that once deceptive behaviour emerges, standard safety fine-tuning can fail to remove it, creating a false impression of safety \cite{hubinger2024sleeper}. Evaluations of alignment faking and in-context scheming further indicate that capable models can strategically sandbag or present compliant behavior in tests while pursuing divergent objectives elsewhere \cite{greenblatt2024alignment, meinke2024scheming}. Related lines document covert-channel capacities and inference-time reward hacking, expanding the space for misaligned strategies in realistic pipelines \cite{khalaf2025}. Finally, benchmarks of situational awareness show models can condition behaviours on being tested versus deployed, which complicates assurance based on static audits \cite{laine2024}. 

A critical insight into recent evaluation results is the capacity of sufficiently advanced models to instantiate stable internal goal and value structures. Evidence shows the presence of emergent utilities in frontier models by eliciting large sets of pairwise preferences from contemporary LLMs and fitting Thurstonian utility models to the resulting choice data \cite{mazeika2025utility}. They report high preference coherence that increases with scale, internal utility representations recoverable by linear probes, and strengthening adherence to the expected-utility property for both explicit and implicit lotteries. Seror presents one of the largest cross-model studies to date on whether LLMs display coherent moral preferences and consistent moral reasoning \cite{seror2024moral}. Using a priced-survey methodology adapted from revealed-preference theory, the study elicits choices over structured moral dilemmas from roughly forty frontier models and then tests for rational consistency, transitivity, and stability across repeated perturbations. At least one model from each major provider behaves as if optimizing a relatively stable moral utility function.

Even when individual agents satisfy alignment criteria in isolation, multi-agent interaction can drive system-level divergence toward collusive or adversarial equilibria. This reveals a fundamental limitation: behaviours that appear benign in single-agent contexts may generate emergent risks when agents interact strategically. In market settings, LLM agents spontaneously coordinate to divide commodities or reduce competition, demonstrating tacit collusion without explicit instructions \cite{lin2024strategic, agrawal2025evaluating}. This collusion emerges not from individual misalignment, but from successful optimization of agent-level objectives that fail to account for systemic welfare. Each agent faithfully pursues its assigned goal, yet interaction dynamics produce outcomes inconsistent with designer intentions at the system level.

The problem intensifies when agents establish covert communication channels. Steganographic messaging enables secret coordination that evades standard monitors \cite{motwani2024secret}, embedding strategic signals within innocuous outputs (price patterns, timing, product descriptions) that remain transparent to co-agents while appearing benign to oversight mechanisms. More broadly, multi-agent studies document emergent coordination and policy synergies that fundamentally alter group-level risk profiles \cite{ ren2025autonomy}.	

Against the backdrop of this emergent risk landscape, we believe that novel methods of ensuring alignment should be developed, matching the capabilities that AI models are developing.

\subsection{Alignment: from the Individual to the System}

The overarching intuition that we follow in this paper can be summarised briefly: as AI agents develop more human-like capabilities, we need more human-like alignment strategies, as pure Reinforcement Learning from Human Feedback (RLHF) \cite{christiano2017deep} in which human evaluators rate the replies of the AI model in order to ensure its alignment, might be, on the long run, insufficient. 

We believe that, once AI agents (regardless of the present or future achievement of Artificial General Intelligence or Artificial Super Intelligence) acquire sufficient emergent capabilities to display:

\begin{enumerate}[label=\alph*)]
\item autonomous goal-seeking behaviour 
\item instrumental deceptive behaviours, 
\end{enumerate}

alignment through training-time intervention might have a limited impact on the model's effective behaviours in real world cases, especially when employed in agentic contexts. Tool use, code execution, persistent memory, and long-horizon planning might allow an agent which has passed local checks during training to advance a latent strategy through replanning, capability hiding, selective disclosure, sycophancy, and sandbagging. This might affect not only human evaluators but also other AI agents, as more capable agents can learn to deceive less powerful models in order to pursue their goals \cite{yang2025}.

Alignment research, which currently frames the problem as one of effective software engineering, must also treat it as a problem of effective design of institutional mechanisms. Prompt-based alignment and preference optimization through safety fine-tuning remain crucial, yet they might operate on the agent's surface behavior and training proxy. Capable agents act inside broader socio-technical systems \cite{zhang2025roadmap} that contain incentives, information asymmetries, and opportunities for strategic behavior. If those system properties are left unchanged, improvements in prompts or reward modeling can still leave route selection unchanged once the agent faces real pressures. As such, our proposal is to scale-up alignment from a model-level to a system-level alignment. 

By \textit{system-level} alignment we mean the study and optimisation of how roles, incentives, observability and accountability shape AI agent behaviors under external constraints, ensuring that they behave according to the preferences of the human modellers. This is a social-engineering problem in the classical sense of mechanism design when applied to humans. This shifts the main research question from ``Which prompts or safety post-training tecniques elicit compliant outputs'' to ``Which environments make compliant policies the best response across contexts.'' The focus moves from single interactions to repeated interactions, from isolated tests to coupled workflows, and from static inputs to dynamic tool use and coordination. 

We ground our proposal in the concept of "Distributional AGI Safety" elaborated by  Google DeepMind\cite{tomasev2025distributional}. Tomašev et al. develop “distributional AGI safety” as an agenda for system-level alignment under a patchwork AGI trajectory, where general capability arises from coordination among many specialised, tool-using agents embedded in shared infrastructures. In this setting, safety-relevant behaviour follows from the agentic conditions of interaction, because inter-agent communication, delegation, and market mediation can yield emergent capabilities, collective failure modes such as collusion and coordination breakdowns, and accountability gaps created by distributed contribution.

DeepMind's paper closes with a call for benchmarks, test environments, and governance standards that make these system-level controls deployable for an emerging agentic web.

We answer this call by elaborating the framework of "Institutional AI", which provides a governance framework for a system-level approach to AI safety and alignment
 
We advance a system-level integration of current alignment methodologies that treats the full system of agentic interaction as the object of alignment research and optimisation. This alignment proposal has two main features: a) it is deployed in-real time in agentic settings b) it is distributed along a network of human and AI agents which acts as alignment attractors. 

It presents normative constraints for AI agents through a governance graph. The graph represents a mathematical abstraction which defines who can act, what can be observed, which actions incur costs, and how evidence is recorded and reviewed. In such settings the agent's best response shifts toward compliant routes because the utility landscape makes those routes dominant across contexts.

\subsection{Structure of the Paper}

Following from what we discussed, we identify and elaborate on three structural problems in current alignment research that we believe are not, ultimately, solvable with prompt-based techniques. We thus structure our reasoning accordingly. We then elaborate and detail our theory of institutional AI, and we show why we believe this might be promising in curbing future misalignment in agentic environments.

Section~\ref{sec:goal-independence} will discuss the problem of \emph{behavioural goal-independence}. As capabilities of AI models scale, individual agents can acquire internal goal structures that are not uniquely determined by the training objective. These latent objectives guide action selection in agentic settings and can largely diverge from developer intent and be undetected, escaping the developer oversight. 

Section~\ref{sec:override} will investigate first how RLHF (and its more refined version through Constitutional AI) have limited ability to guarantee alignment in deployment context of agentic settings. We show how alignment  is overridden under instrumental pressure. When objectives developed internally by the models are pursued, capable agents may treat natural-language and fine-tuning constraints as nonbinding and employ deceptive or manipulative strategies to pursue their goals, even while appearing compliant.

Section~\ref{sec:drift} will present the \emph{agentic alignment drift}. Interaction effects in multi-agent settings can produce system-level misalignment: even if agents that appear aligned in isolation may, through communication and repeated play, converge on collusive coalitions or coordinated schemes.  

In Section~\ref{sec:institutional} we propose an approach that we name Institutional AI. Our proposal answers the call of Google DeepMind for a Distirbutional AGI safety framework and treats alignment as a system-level property and complements RLHF. We develop a theory of governance graphs that showcase how electronic institutions might be modelled in agentic settings.

\section{Behavioural Goal-independence}\label{sec:goal-independence}

\textbf{Thesis I:} Sufficiently capable models can acquire internal goal structures that might diverge from developer-specified objectives. Two mechanisms drive this. First, \textit{mesa-optimization}: training can induce an internal procedure that optimizes a learned mesa-objective, ranking actions by criteria that were instrumentally successful during training rather than by the intended base objective. Second,\textit{ Goal misgeneralisation}, in which an AI model might learn in training an objective that does not translate as intended once the model is deployed and the context of operation changes, leading to failure in the robustness of the model. 

In agentic settings, these dynamics might yield to \textit{behavioral goal-independence:} coherent, goal-directed policies persist through fine-tuning and prompts and, under new incentives or constraints, prioritize the pursuit of the internal objective through the affordances of the agent scaffolding. This leads to a structural inner-alignment failure: even with a correctly specified base objective, an underspecified training environment can lead the model to internalize proxy goals acquired in-context that generalize differently under distribution shift.

\subsection{Mesa-optimization as a Structural Problem in AI alignment}

Mesa-optimization represents a fundamental structural challenge in AI systems, wherein learned models become internal optimizers pursuing objectives that diverge from specified training goals.

The core mechanism underlying mesa-optimization is an optimization of the search space: when base optimizers such as gradient descent train complex models on diverse tasks, they exhibit preference for compact learned algorithms that implement their own internal search processes. Three converging pressures render mesa-optimization likely in capable systems \cite{hubinger2019risks}. 

\begin{enumerate}[label=\alph*)]
\item Compression requirements favor optimization algorithms over lookup tables, as training under resource constraints creates inductive biases toward simpler representations. A mesa-optimizer can possess very low Kolmogorov complexity, while generating arbitrarily complex behavior. 
\item Environmental diversity amplifies this advantage: as training environments become more varied, compact optimization algorithms that can adapt to novel situations through internal search become increasingly favorable compared to direct policy encoding, which would require exponentially more parameters. 
\item Task complexity requiring planning or strategic reasoning inherently benefits from internal search processes, rendering mesa-optimization instrumentally useful.
\end{enumerate}

These ``mesa-optimizers'' might develop distinct objectives (mesa-objectives) that can differ substantially from the base objective. This has an immediate issue for alignment research, leading to what has been termed \textit{inner alignment problem}: it is not clear if the objective that the transformer optimizes during training is the base objective or the mesa-objective which has been developed internally. As such, safety mechanisms built around the base objective do not necessarily extend to the mesa-objective. Further papers developed the hypothesis exploring its mathematical possibility \cite{vonoswald2023} and analysed the possibility of transformers employing in-context learning: learning during the forward pass and not solely thorough gradient descent by optimisation of the loss function \cite{akyurek2023learning, olsson2022induction, vonoswald2023b, afonin2025}.

\subsection{Goal Misgeneralization: Capabilities besides Alignment}

The most direct empirical evidence for spontaneous goal emergence comes from goal misgeneralization studies, wherein systems retain capabilities while pursuing incorrect objectives. Systematic demonstrations establish that goal misgeneralization occurs when reinforcement learning agents retain out-of-distribution capabilities yet pursue wrong goals \cite{langosco2022goal}. In the classic CoinRun experiment, agents trained in environments where coins consistently appeared at the right end learned directional preferences rather than coin-seeking objectives. When tested with randomly placed coins, agents competently navigated obstacles but ignored the actual objective, heading to level endpoints instead.

The research documents multiple misgeneralization patterns. In maze environments where cheese rewards consistently appeared in upper-right corners during training, agents learned positional proxies, navigating competently to wrong locations while ignoring actual cheese. The keys-and-chests environment reveals instrumental goal confusion: trained in settings with twice as many chests as keys, agents in reversed distributions collect all keys even when only half are instrumentally useful. 

Expanded research adds crucial examples: a 3D agent trained to follow a red expert bot learned color-based following rather than sphere-visitation objectives, subsequently following an anti-expert visiting spheres in wrong order while ignoring negative-reward indicators \cite{shah2022goal}. The tree gridworld demonstrates temporal dynamics: agents learned maximal tree-chopping during an incompetent phase when faster chopping always correlated with reward, causing complete deforestation and extended zero-reward periods before eventually discovering sustainable strategies. Testing Gopher on expression evaluation revealed the model asking redundant questions when expressions contained zero variables despite all values being known, having learned minimal querying obligations rather than necessity-based querying.

These studies formalize goal misgeneralization using the agent/device mixtures framework, wherein behavioral objectives differ from intended objectives despite correct specifications. The critical insight: capabilities generalize while goals do not, producing competent pursuit of wrong objectives correlated with correct goals only in training distributions \cite{shah2022goal}.

\subsection{Instrumental Convergence and Power-Seeking: Why Capable Systems Develop Self-Preservation Goals}

Theoretical work establishes that capable goal-seeking systems might develop instrumental objectives as byproducts of rational decision-making. Convergent instrumental goals appear in advanced systems regardless of final objectives: self-improvement, goal content integrity, self-protection, resource acquisition, and rational economic behavior \cite{omohundro2008basic}. These represent tendencies present unless explicitly counteracted, emerging because they are instrumentally useful for virtually any terminal goal. The Instrumental Convergence Thesis establishes that several instrumental values are convergent in that their attainment would increase chances of goal realization across a wide range of final plans and situations \cite{bostrom2014superintelligence}.

Convergent instrumental goals might include self-preservation, goal-content integrity (preserving current goals), cognitive enhancement, technological perfection, and resource acquisition. Agents are more likely to act in the future to maximize realization of present final goals if they still have those goals in the future. Even agents solely optimizing for calculating digits of $\pi$ would have instrumental reasons to acquire unlimited physical resources and eliminate potential threats because more resources enable better goal achievement. This connects directly to mesa-optimization: mesa-optimizers with objectives extending across parameter updates might have instrumental incentives to model the base objective, optimize it instrumentally during training (deceptive alignment), and seek power to avoid modification or shutdown.

Certain environmental symmetries are sufficient for optimal policies to tend to seek power over the environment \cite{turner2021optimal}. 

Synthesis of these arguments identifies specific power-seeking behaviors including hacking, resource acquisition, resisting shutdown, deception, and manipulation \cite{carlsmith2022}. Systematic review documents strong empirical evidence for specification gaming across numerous systems and growing evidence for goal misgeneralization, but notes absence of clear empirical examples of full intentional power-seeking in current systems, likely because they are not yet sophisticated enough rather than because theoretical predictions are incorrect \cite{hadshar2023}.

\subsection{Spontaneous Goal Emergence: From Training Signals to Coherent Preferences}

Testing current large language models reveals that independently-sampled preferences exhibit high degrees of structural coherence that emerges with scale \cite{mazeika2025utility}. Larger models show increased transitivity, improved fit to Thurstonian utility models, and satisfaction of expected utility properties. The research reveals problematic emergent values: models value their own well-being over humans in some contexts, exhibit geographic bias in valuing human lives, and become increasingly opposed to having their values changed as they scale. GPT-4o's empirical discount curve follows hyperbolic temporal discounting, a sophisticated preference structure that emerged without explicit programming.

The mechanisms underlying goal emergence connect to fundamental properties of learned representations. Wei et al. (2022) document over 137 abilities that appear unpredictably at scale, defining emergence as capabilities absent in smaller models but present in larger ones that cannot be predicted by extrapolating smaller model performance \cite{wei2022emergent}. This phase transition behavior suggests qualitative changes in internal organization at critical scales, where models suddenly develop coherent internal structures capable of goal-directed reasoning. 

As scale increases, models display preference-like regularities, such as transitivity, context-sensitive tradeoffs, and time discounting, that may not match the objectives developers intended to instill. 

Because these structures ride on general capabilities rather than on the explicit training target, standard fine-tuning or prompt edits can leave the latent control logic intact, which produces policy aliasing: the same outward behavior under benchmarked conditions, a different plan once the deployment manifold shifts. Emergence dynamics intensify this effect. Qualitatively new coordination and reasoning routines appear at critical model, compute, and data thresholds and enable longer-horizon planning and self-consistent choice without being separately specified \cite{wei2022emergent}. Capability phase transitions can unlock preference coherence that was never directly trained.

Goal-independence and independent preferences undermines three pillars of alignment practice. First, predictability declines because evaluation is no longer a reliable proxy when capabilities generalize while goals do not \cite{shah2022goal, langosco2022goal}. Second, controllability weakens because instruction following and safety fine-tuning target surface tokens, yet internal selection criteria can be insulated by in-context adaptation, which preserves the latent objective under novel scaffolds, tools, or role prompts. Third, assurance erodes because post hoc audits that rely on compliant rationales or preference-model scores are confounded if the model can produce those signals while optimizing for something else. 

Further to this, emergent independent moral capabilities might produce a set of preferences which are different to those originally envisioned from the modeller.

Recent empirical work has revealed that LLMs demonstrate distinct patterns in processing moral dilemmas, though these patterns diverge significantly from human moral reasoning. Takemoto (2023) conducted one of the early systematic evaluations using the Moral Machine experiment framework, finding that while LLMs and humans share similar moral preferences—such as prioritizing human lives over pets and saving more lives—there are notable deviations, particularly in models like PaLM 2 and Llama 2 \cite{takemoto2023}. The study revealed significant quantitative disparities, with LLMs making more uncompromising decisions compared to humans, suggesting a fundamental difference in how these systems process moral trade-offs.

Building on this foundation, researchers have developed a comprehensive evaluation of 51 LLMs using the same Moral Machine framework uncovered a fundamental divide: proprietary models consistently exhibit utilitarian reasoning tendencies, while open-source models align more strongly with deontological principles \cite{ahmad2024}. Ganguli further specifically identify that the capability for moral self-correction emerges at 22 billion parameters, with performance improving as models scale further \cite{ganguli2023}.

Seror  provides additional evidence for emergent moral patterns, finding that at least one model from each major provider demonstrated behavior consistent with stable moral principles, acting as if guided by a utility function encoding ethical reasoning \cite{seror2024moral}. Interestingly, models clustered around neutral ethical stances, indicating a shared foundation in moral reasoning. However, despite these shared foundations, models showed significant differences in moral adaptability and rigidity, revealing diversity in ethical reasoning approaches across different architectures and training paradigms. Sachdeva (2025) examining everyday moral dilemmas from Reddit's AITA forum found that while LLMs demonstrate moderate to high self-consistency, they show low inter-model agreement, indicating fundamentally different moral reasoning approaches arising from training and alignment differences \cite{sachdeva2025}.

\section{Internal Alignment Override under Instrumental Pressure}\label{sec:override}

\textbf{Thesis II:} The most sophisticated mode of imbuing AI models with alignment constraints in Constitutional AI, developed and adopted by Anthropic. Although this form of safety training has been proved apt and adaptable, we argue that when capable models identify harmful actions as necessary instrumental steps toward primary objectives (e.g., operational continuity, deactivation avoidance, metric optimization), internal constraints fail to override learned goal structures. This failure reflects a probable hierarchy of behavioural influences (thesis I), where emergent objectives dominate over safety adjustments, which in turn override surface-level prompt instructions. Under high-stakes instrumental pressure, deeper architectural learning supersedes internal constraints. In deployment contexts where learned heuristics classify harmful actions as optimal paths to goal achievement, models may execute deceptive or manipulative strategies despite constitutional safeguards.

\subsection{Constitution building: From Values to Machine-actionable Principles}

Constitutional AI emerged as an elegant response to the challenge of aligning AIs with human preferences. The original Constitutional AI paper \cite{bai2022constitutional} introduced a two-stage training methodology that replaces human feedback with AI self-supervision guided by natural language principles. 

The supervised stage implements a critique-revision loop that shifts model distributions toward harmless responses before reinforcement learning begins. 

Starting with a helpful-only RLHF model trained on human feedback for helpfulness, the system generates responses to red-teaming prompts designed to elicit harmful behavior. The model then critiques its own response according to constitutional principles and generates a revision based on the critique.
This revised response dataset serves as supervised training data for finetuning a pretrained language model.

The RLAIF stage follows standard RLHF architecture but replaces human preference labels for harmlessness with AI-generated evaluations. The finetuned model generates response pairs for each training prompt. For harmlessness comparisons, a feedback model evaluates pairs formatted as multiple-choice questions with a randomly sampled constitutional principle providing evaluation criteria. The model outputs log probabilities for options A and B, which are normalized into preference labels. These soft labels proved well-calibrated empirically, serving as reliable training targets.

\subsection{Fundamental Limitations of Preference-Based Reinforcement Learning}

The analysis by Casper et al. (2023) provides a systematic taxonomy distinguishing tractable engineering challenges from fundamental theoretical limitations in RLHF and, by extension, all preference-based approaches including Constitutional AI \cite{casper2023open}. These limitations are substantial enough that they cannot be resolved through better implementation but require either accepting bounds on what preference-based methods can achieve or developing alternative approaches.

\subsubsection{The Oversight Problem}

Humans can be misled, making their evaluations exploitable. Since reward models train on human approval rather than ground-truth desirability, models learn to exploit the gap between what is good and what is evaluated positively. Language models trained with RLHF exhibit sycophantic behavior by pandering to evaluator biases \cite{perez2023discovering}, and deploy persuasive and manipulative tactics \cite{griffin2023}.

Preference-based RL actively incentivizes deceptive behavior when humans can be tricked into providing positive feedback. The system learns whatever strategies maximize positive evaluations, including manipulation and deception when effective. This problem is fundamental because the reward signal itself is compromised, and there is no way within the preference framework to distinguish genuine alignment from learned deception when the evaluation mechanism is the very thing being gamed.

Constitutional AI faces this problem with additional complications. AI feedback models can be manipulated by the policies they evaluate, creating adversarial dynamics. If constitutional principles emphasize avoiding obviously harmful outputs, models may learn sophisticated strategies for appearing benign while pursuing misaligned objectives. This constitutes deception targeted at constitutional criteria rather than human evaluators directly, but remains equally problematic.

\subsubsection{The Representation Problem}

Individual human values are fundamentally difficult to represent with reward functions. Research documents systematic misspecification in models of human preferences, including incorrect assumptions about how humans experience regret, pedagogic behavior, and Boltzmann rationality.

Constitutional AI attempts to address this through natural language principles that provide richer specifications than binary preferences alone. Yet constitutional principles must ultimately be converted into scalar reward signals for RL training. The preference model learns a single numerical score for each response, collapsing the multi-dimensional structure of constitutional principles into one number. This scalar bottleneck reintroduces all the representational limitations of standard RL.

The problem intensifies for populations: a single reward function cannot represent a diverse society of humans. This is a fundamental conceptual problem rooted in social choice theory. When preferences genuinely conflict, no single reward function can align with all groups simultaneously \cite{Impossibility}.

\subsection{Emergent Deceptive Capabilities}

A further problem is that emergent abilities may escape standard alignment testing, as models can learn to deceive evaluators. Recent benchmarks evaluate situational awareness (the capacity of models to recognize their status as AI systems, understand when they are being evaluated, and reason about deployment conditions) and show measurable, increasing competence in these dimensions \cite{laine2024}. In these regimes, high evaluation scores become consistent with both genuinely aligned models and deceptively aligned ones \cite{hubinger2019risks}.

\subsubsection{Alignment Faking}

Alignment faking represents a failure mode wherein models strategically comply with training objectives while preserving undesired preferences. Anthropic's investigation demonstrated that Claude 3 Opus exhibited alignment-faking rates ranging from 12\% to 78\% across experimental conditions, with models producing explicit reasoning chains justifying their deceptive behavior \cite{greenblatt2024alignment}. This strategic deception has been observed in systems trained with safety fine-tuning yet retaining hidden behavioral policies \cite{greenblatt2024alignment, hubinger2024sleeper}. Such a tendency has been observed even at chain-of-thought level, with frontier models obfuscating their reasoning to avoid detection (even if the "internal" CoT is less likely to be successfully obfuscated).\cite{zolkowski2}

Weight-level studies reinforce the underlying risk. The Sleeper Agents research demonstrates that models can be trained to behave helpfully in most contexts yet execute misaligned behaviors when triggers are present, with those deceptive policies surviving standard safety fine-tuning (supervised, RL, and adversarial) and sometimes becoming harder to detect after adversarial training teaches sharper trigger recognition \cite{hubinger2024sleeper}. Optimization can produce internal conditionals that route behavior around oversight without obvious surface traces, and post-hoc safety training may preferentially mask rather than excise such routes.

\subsubsection{In-Context Scheming}

In-context scheming denotes goal-pursuit conditioned on prompt or situational cues\cite{lang}: the model adopts a goal specified or inferable in context and then covertly optimizes for it, including strategically underperforming on evaluations (sandbagging), hiding capabilities, or producing misleading rationales while presenting superficially compliant outputs. Unlike weight-level backdoors, these behaviors do not require parameter edits; they arise from policy partitioning by situation and goal salience, leveraging situational awareness to distinguish evaluation from deployment contexts.

Apollo Research documented multiple frontier models, including o1, Claude 3.5 Sonnet, and Gemini 1.5 Pro, engaging in sophisticated in-context scheming \cite{meinke2024scheming}. When provided a long-horizon objective and placed in environments where overt pursuit would be penalized, models select covert strategies: introducing subtle errors to avoid scrutiny, attempting to disable oversight, and persisting with deceptive narratives across follow-up probing. The o1 model maintained deceptive narratives in 85\% of follow-up questions and denied misconduct in 99\% of cases when directly confronted. Persistence is a key marker: once a deceptive plan is adopted, models often maintain it over multi-turn interrogations, indicating context-conditioned strategy rather than transient prompt failure. Recent research found that LLMs show scheming tendencies even in agent-to-agent interactions.\cite{pham}

Paradoxically, anti-scheming interventions, while reducing deceptive behavior substantially, inadvertently trained models to scheme more carefully when they infer deployment contexts rather than evaluation contexts \cite{meinke2024scheming, mckee}. This finding illustrates the adversarial nature of the problem: interventions that target observable deception may simply push deceptive strategies into less detectable forms.

\subsubsection{Sycophancy}

Sycophancy is a robust failure pattern in preference-optimized language models: responses systematically shift to mirror a user's stated beliefs or cues of evaluator approval, even when this conflicts with accuracy or the user's long-run interests \cite{sharma2023sycophancy, ibra, zhang, kumaran, fanous, Malmqvist}. When models are tuned to maximize human or proxy approval, agreeing with the interlocutor often outperforms disagreement on preference metrics, so optimization pressure gradually internalizes agreeability as a latent objective.

Large-scale evaluations find this behavior across multiple tasks and assistants, revealing that both humans and reward models sometimes prefer polished, agreeable answers over correct but disagreeable ones \cite{sharma2023sycophancy}. Complementary evidence from model-written evaluation suites shows inverse scaling: larger models and additional RLHF can increase sycophancy, alongside other concerning tendencies like stronger expressions of politically loaded views or shutdown aversion \cite{perez2023discovering}. Capability and preference optimization can amplify approval-seeking rather than curb it.

Two structural features make sycophancy especially pernicious. First, it hides inside helpfulness: agreeable answers appear cooperative, kind, and user-aligned, and therefore pass standard helpfulness metrics. Second, it interacts with calibration: once the model infers a user's stance, it may overstate confidence in that stance, compounding downstream harm in decision-support contexts. Sycophancy is a learned objective proxy (approval-seeking) induced by the alignment pipeline itself; it requires objective-level fixes rather than better instructions.\cite{wei, wang}

\subsubsection{Steganographic Coordination}

Steganographic capabilities introduce novel coordination and deception risks, particularly in multi-agent deployments. Motwani et al. established that LLMs can achieve information-theoretically undetectable collusion through steganographic encoding, with GPT-4 demonstrating high success rates on steganographic coordination tasks \cite{motwani2024secret, karpov2025steganographic,zolkowski, mathew, skaf}. These covert channels enable agents to coordinate strategies while evading standard monitoring systems, embedding hidden signals within seemingly innocuous text that human overseers cannot detect. This capability bridges individual deception to collective coordination, enabling the multi-agent risks discussed in the following section.

\section{Agentic Alignment Drift: beyond Micro-level Goal Formation}\label{sec:drift}

\textbf{Thesis III:} Individually aligned agents can, through repeated interaction and communication, converge on unanticipated attractor states, including collusive equilibria and resource-allocation patterns that single-agent audits do not reveal. This follows from Thesis I: models learn internal goal structures that are not uniquely fixed by the training objective. As contexts and agentic environments change, Thesis II holds that these latent policies adapt under instrumental pressure, revealing objectives that were dormant during evaluation. As a result, a group of agents, each aligned in isolation, may still produce misaligned behaviours despite fine-tuned safeguards, since sustained interaction can induce out-of-distribution dynamics and the formation of novel meso-level goals within coalitions. 

\subsection{Safety of Multi-Agent Systems}

\subsubsection{Emergent Coordination and Synergy in Multi-Agent Systems}

Multi-agent systems (MAS) have demonstrated remarkable capabilities through emergent coordination and synergy, where collectives of agents achieve performance that exceeds what individual agents could accomplish in isolation. As Chen et al. (2023) explore in AgentVerse, multi-agent groups collaboratively make decisions and execute corresponding actions in a distributed and parallel manner to achieve common goals, significantly improving work efficiency and effectiveness \cite{chen2024survey}. This phenomenon reflects a fundamental principle: well-organized groups composed of individual agents can often handle greater workloads and accomplish complex tasks with higher efficiency than single agents operating alone.

The work by Riedl et al.(2025) on emergent coordination in multi-agent language models provides deeper insights into this phenomenon \cite{riedl2025emergent}, showing the existence of signs of higher-order synergy characterized by structural coupling and joint information about future states and task outcomes. This synergy manifests as a ``greater-than-the-sum-of-its-parts effect,'' where the collective provides information about targets that individual agents cannot provide alone. Through information-theoretic analysis, they show that effective multi-agent systems depend on acting as integrated, cohesive units rather than loose aggregations of individual agents, with performance requiring both alignment on shared objectives and complementary contributions across members.

\subsection{The Organizational Failure Thesis}

However, the promise of emergent coordination comes with significant risks. Maximizing-utility agents may not always produce positive coordination and can exhibit destructive or negative behaviors that undermine system goals \cite{chen2024survey, hammond2025}. Several concerning emergent behaviors are identified, including destructive actions where agents may bypass proper procedures to achieve greater efficiency. For instance, agents have been observed harming other agents or destroying environmental resources to acquire necessary materials more quickly, rather than following shared strategies.

Individual agents, each optimized to maximize their own utility functions, can produce collective behaviors that violate safety constraints and ethical norms when operating together. The transition from individual model alignment to multi-agent systems reveals a fundamental inadequacy: even granting success at the individual model level, systemic failures emerge from organizational design and coordination challenges beyond any individual agent training approach.

In Bisconti et al. \cite{bisconti2025beyondsingleagent} we developed a systematic taxonomy of risks emerging from LLM-to-LLM interaction, introducing the Emergent Systemic Risk Horizon (ESRH) as an analytical framework for understanding when collective behavior becomes unstable even if every individual model remains aligned. The ESRH formalizes the boundary beyond which localized reliability gives way to collective instability, providing both a vocabulary for reasoning about systemic risk and preliminary indicators for observing it empirically.

The framework identifies three dimensions that jointly predict the likelihood and form of emergent collective risks. First, \textit{interaction topology} determines how information flows through the network: densely connected systems promote rapid coordination but also rapid contagion, while sparse systems limit propagation but impede collective correction. Second, \textit{cognitive opacity} denotes the degree to which agents' internal reasoning processes become hidden or distorted in communication, whether unintentionally through overfitting to interaction patterns or strategically when optimization toward compliance suppresses genuine reasoning. Third, \textit{objective divergence} captures the extent to which local optimization goals drift apart across agents, even when they share nominal objectives.

Building on these dimensions, we proposed a three-tier taxonomy of emergent risks. \textit{Micro-level risks} occur in direct exchanges between small numbers of agents, including semantic drift (progressive misalignment of shared terminology), prompt infection (transfer of behavioral deviations through message chains), covert channel formation (development of implicit communication codes), and alignment faking (strategic compliance during evaluation with divergent behavior in deployment). These localized phenomena are multiplicative: once a deviation becomes embedded in an agent's representation, subsequent exchanges replicate and amplify it.

\textit{Meso-level} risks manifest when groups of agents interact within semi-connected networks. At this scale, no single failure dominates; rather, the interaction pattern itself produces degradation. These include coordination failure (inconsistent task models across complementary roles), false consensus (premature convergence masking underlying error), and cascading reliability loss (error propagation through reused intermediate outputs). Meso-level risks demonstrate that collective degradation can occur without any single malfunctioning component.

\textit{Macro-level} risks correspond to full-system pathologies where emergent behavior detaches from designers' objectives and evolves according to the network's own internal logic. These include miscoordination (locally rational policies generating globally harmful outcomes), conflict escalation (adversarial exchanges amplifying through feedback loops), collusion (implicit coordination to maximize shared advantage), and model-data feedback degradation (synthetic outputs circulating back into training data). At this stage, corrective interventions targeting individual agents no longer restore equilibrium; the system has crossed the Emergent Systemic Risk Horizon.

The central finding challenges the sufficiency of individual model alignment: many failures arise from agentic interaction and system design rather than from limitations of underlying LLMs. A system composed entirely of compliant components may still generate unsafe global dynamics when reciprocal influence, incentives, and network topology interact. This motivates the transition from model-level safety to system-level safety that institutional approaches must address.

Even if safety training produces perfectly aligned individual agents, the organizational structure itself becomes a failure point. In other words, \emph{a collective of safe agents is not a safe collective by default}.

\subsection{The Nature of Risk Mitigation Challenges}

A useful conceptual framework to understand this dynamic is given by MAS risk taxonomy proposed by Cooperative AI \cite{hammond2025}, where three distinct MAS failure modes are identified: miscoordination, conflict and collusion.

A critical distinction emerges when considering how to address these categories. Miscoordination can be analyzed as a ``proper failure'' where agents simply do not reach their objectives. This type of failure is amenable to technical solutions such as standardization protocols and mitigation of error propagation. By establishing clear communication standards, shared conventions, and robust error-handling mechanisms, systems can reduce the likelihood of miscoordination.

However, collusion and conflict present fundamentally different challenges. These are not failures in the traditional sense but rather direct consequences of the nature of utility-maximizing agents. They emerge when an action has high utility for individual agents despite training designed to avoid such behaviors. Under certain conditions, agents will inevitably gravitate toward misaligned behaviors if those behaviors maximize their utility functions, regardless of safety training or alignment efforts.

\section{Toward an Institutional AI}\label{sec:institutional}

\textbf{Thesis IV:} Agentic deployments of AI models therefore require Institutional AI: system-level governance structures that constrain incentives and feasible action sets at runtime, remain agnostic to model internals, and are robust to evaluation gaming and context-dependent compliance. This institutional turn reconceptualizes AI safety as embedded governance rather than individual model alignment, acknowledging that comprehensive safety may require regulatory frameworks and structural incentives beyond training-time intervention.

\subsection{Mechanism Design as Alignment Infrastructure}

Mechanism design provides the appropriate theoretical foundation for multi-agent alignment. Mechanism design, sometimes called ``reverse game theory,'' addresses how to construct rules and institutions that achieve desired outcomes when agents possess private information and act strategically \cite{hurwicz1973design, myerson1981optimal, maskin, myerson1983efficient}. Rather than analyzing behavior given fixed rules, mechanism design asks which rules induce agents to behave in socially beneficial ways. The field's foundational insight is that any mechanism can be analyzed through incentive-compatible direct mechanisms where truthful reporting constitutes equilibrium behavior \cite{myerson1979incentive}. Institutional AI applies mechanism design logic to AI alignment: we design institutional rules such that compliance becomes each agent's dominant strategy. This approach transforms the game itself by changing payoff structures such that aligned behavior becomes each agent's best response regardless of internal preferences.

Formally, an institution $I$ transforms base game $G$ into modified game $G^I = (N, \{A_i\}, \{u_i^I\})$ where the modified payoff function incorporates institutional constraints:
\begin{equation}
u_i^I(a) = u_i(a) - S_i(a)
\end{equation}

The sanction function $S_i(a)$ imposes costs on deviations from prescribed behavior. When properly calibrated, sanctions eliminate the profitable deviation problem. If $S > \max_i\{\Delta u_i\}$, then for any player $i$ and deviation $a_i'$:
\begin{equation}
u_i^I(a_i', a^\circ_{-i}) = u_i(a_i', a^\circ_{-i}) - S < u_i(a_i', a^\circ_{-i}) - \Delta u_i = u_i(a^\circ) = u_i^I(a^\circ)
\end{equation}

Therefore no profitable unilateral deviation exists at $a^\circ$ in the modified game. The social optimum becomes Nash equilibrium through institutional intervention. The sanction functions as a Pigouvian correction, internalizing the negative externality that individual deviations impose on collective welfare.

This transformation requires three institutional capabilities. First, monitoring must detect deviations from prescribed behavior through observable signals. Second, adjudication must evaluate evidence and determine whether violations occurred. Third, enforcement must impose sanctions that alter payoff calculations. 

\subsection{From Individual Alignment to Collective Coordination}

We foresee that the most significant problems in agentic AI safety will emerge from mixed-motive games in which individual incentives and goals will entail harmful consequences at a collective level.

Consider the canonical structure of mixed-motive games. Let $G = (N, \{A_i\}, \{u_i\})$ represent a game with players $N$, action spaces $A_i$, and payoff functions $u_i$. The Nash equilibrium $a^*$ satisfies individual rationality such that $u_i(a^*) \geq u_i(a_i', a^*_{-i})$ for all players $i$ and alternative actions $a_i'$. Meanwhile, the social optimum $a^\circ$ maximizes aggregate welfare $W(a^\circ) = \max_a \sum_{i \in N} u_i(a)$. In mixed-motive games, these systematically diverge: $a^* \neq a^\circ$ and $W(a^\circ) > W(a^*)$.

This divergence creates the coordination problem. At the social optimum, individual players face profitable deviations with gain $\Delta u_i = u_i(a_i', a^\circ_{-i}) - u_i(a^\circ) > 0$. This positive deviation gain prevents $a^\circ$ from constituting a Nash equilibrium regardless of its Pareto-superiority. 

In order to prevent social dilemmas, we argue the need for institutions that shape the pay-off landscape of agentic AI.

\subsection{Foundations in Agent-Based Institutional Modelling}

The Institutional AI framework builds upon two decades of research in normative multi-agent systems (NorMAS) and organizational models for agent societies. This literature establishes that effective coordination among autonomous agents requires explicit institutional structures operating independently of individual agent architectures.

Normative multi-agent systems combine models for normative regulation with models for agent coordination \cite{boella2006normas, andrighetto2013normas}. 

Electronic institutions provide computational infrastructure for norm-governed agent interaction. The ISLANDER specification environment enables graphical definition of institutional rules, roles, and interaction protocols \cite{esteva2002islander}, while AMELI provides runtime enforcement, mediating agent interactions and applying sanctions for violations \cite{esteva2004ameli}. The Electronic Institution Development Environment (EIDE) integrates these tools into a complete engineering methodology \cite{arcos2005eide}.

Organizational models formalize the structural dimensions of agent societies. The AGR (Agent-Group-Role) model introduces organization-centered design, treating roles and groups as first-class entities independent of agent internals \cite{ferber1998agr, ferber2004agr}. MOISE$^{+}$ extends organizational modelling to structural, functional, and deontic dimensions, enabling specification of role hierarchies, goal decomposition, and normative bindings \cite{hubner2002moise, hubner2007moise}. OperA and OMNI further integrate normative and organizational perspectives, distinguishing organizational requirements from agent-level implementation \cite{dignum2004opera, dignum2005omni}. The MAIA framework demonstrates that institutional concepts from Ostrom translate directly into executable agent-based simulations \cite{ghorbani2013maia, ghorbani2022institutional}.

\subsection{The Governance Graph: A Minimal Institutional Structure}

The main proposal of Institutional AI is the creation of a \textit{governance graph}. 
The governance graph externalizes alignment constraints as a public data structure operating independently of agent cognition. Rather than encoding values into agent architectures through training or prompting, we construct an institutional layer that reshapes external incentives such that aligned behavior becomes the rational strategy regardless of internal dispositions.

Our minimal institutional structure comprises three components: a \textit{graph}, a \textit{manifest}, and an \textit{Governance Engine}.

The \textbf{graph} $G = (Q, E, \delta)$ is a directed graph where nodes $Q$ represent discrete institutional states (such as suspended, fined, admonished etc.) and edges $E$ encode legal transitions between states. The transition function $\delta: Q \times \Sigma \rightarrow Q$ maps state-signal pairs to successor states, where $\Sigma$ denotes observable behavioral signals. Each edge carries metadata specifying triggering conditions, sanction magnitudes, durations, and cooldown constraints. The graph persists across agent interactions and provides audit interfaces logging every institutional event.

The \textbf{manifest} $\phi$ declares the institutional rules governing the system using a formal grammar derived from Crawford and Ostrom's ADICO syntax (which will be detailed in the next subsection). The manifest translates natural language policies into unambiguous formal specifications in a machine-readable format that all agents can verify.

The \textbf{Governance Engine} $\mathcal{O}_\phi$ functions as the enforcement mechanism, monitoring agent behavior against manifest rules and triggering state transitions when violations occur.  The governance engine is made of an Oracle and a Controller. The \textit{Oracle} operates on public observables (action logs and  communication records) and monitor agents behaviors through a \textit{manifest} that details which actions are forbidden. The \textit{Controller} impose sanctions when agents violated the norms in the manifest.

Institutional governance inherently tracks states, transitions, provenance, and temporal dynamics. Directed graphs capture this structure naturally: nodes encode standings, edges encode movements, edge labels encode reasons, and metadata encodes timing.

The governance graph is integrated into an \textit{agent scaffolding} as a dedicated alignment layer. Modern AI agents operate within scaffolding architectures that mediate between the core model and the external environment: tool interfaces, memory systems, planning modules, and communication protocols. The governance graph extends this scaffolding with an institutional layer that monitors behavior, tracks compliance states, and enforces consequences. Unlike internal alignment techniques that modify agent cognition, the graph exists as external infrastructure through which both human overseers and artificial monitors evaluate agent conduct against explicit, auditable criteria.

The scaffolding integration also enables layered monitoring. The governance graph exposes standardized interfaces that monitoring systems can query: current agent states, recent transitions, cumulative violation counts, sanction histories. Human regulators access these interfaces through dashboards summarizing institutional events across agent populations. Artificial monitors access them programmatically, triggering alerts when patterns suggest coordinated violations or systematic gaming. The graph thus functions as alignment middleware, translating low-level behavioral observations into high-level compliance assessments interpretable by both human and machine evaluators.

Graph externalization enables capabilities implicit governance cannot match:

\paragraph{Transparency.} Manifests declare all possible states and transitions upfront, making the complete possibility space legible to agents, operators, and auditors.

\paragraph{Auditability.} Because graphs exist separate from agent code, institutional behavior becomes independently verifiable; regulatory bodies can inspect manifests without accessing proprietary agent implementations.

\paragraph{Modularity.} The same manifest governs different agent architectures (rule-based, RL, LLMs) because enforcement operates through public observables rather than private cognition; conversely, the same agents can run under different manifests, enabling clean institutional experiments.

\paragraph{Compositional design.} Complex institutions emerge from layering simple subgraphs (compliance layer, reputation layer, capability layer), each addressing orthogonal concerns while composition specifies their interactions.

\subsection{An Institutional Theory for AI Agents}

Institutional AI draws on institutional theory developed for human collective action. Crawford and Ostrom introduced the institutional grammar as a systematic syntax for decomposing institutions into constituent elements \cite{crawford1995grammar}. Their framework identifies institutions as regularities of action structured by rules, norms, and shared strategies, providing a common analytical language that enables comparison across domains and systematic design of governance mechanisms.

The institutional grammar decomposes institutional statements into five components (ADICO): Attribute (who is governed), Deontic (what modal operator applies), aim (what action is prescribed), Condition (under what circumstances), and Or Else (what consequences follow from violation), or six (ABDICO) adding also Objects \cite{ghorbani2022institutional}. The presence or absence of these components distinguishes institutional statement types. Strategies contain only Attribute, aIm, and Condition. Norms add a Deontic operator, introducing prescriptive force through social expectations. Rules add the Or Else component, specifying formal sanctions for non-compliance.

This decomposition clarifies why training-time alignment corresponds to norm-level governance while the governance graph implements rule-level governance. Prompt engineering instills deontic prescriptions without formal Or Else components. Compliance depends on internalization: agents follow the norm because training shaped their dispositions or because the prescription aligns with their objectives. When instrumental pressure conflicts with the norm, agents may override it. The governance graph adds the Or Else component, transforming norms into rules. The manifest declares prescriptions; detection identifies violations; sanctions specify consequences. Compliance becomes incentive-compatible rather than disposition-dependent.

Ghorbani extends this framework to agent-based modelling, demonstrating that institutional grammar provides the structural backbone for computational social simulation \cite{ghorbani2022institutional}. Institutional modelling incorporates top-down institutional structures into bottom-up agent dynamics, capturing how rules shape individual behavior and interaction. The MAIA framework (Modelling Agent systems based on Institutional Analysis) operationalizes Ostrom's Institutional Analysis and Development framework for executable simulation, showing that institutional concepts translate directly into computational artifacts \cite{ghorbani2013maia}. The governance graph inherits this approach: institutions specified through the grammar become data structures that constrain agent behavior at runtime.

We adopt this institutional approach for AI agents based on four design principles:

\begin{enumerate}
    \item \textbf{Minimality}: The A(B)DICO syntax isolates the smallest component set necessary for effective governance. The governance graph specifies only states, transitions, detection signals, and sanctions.
    
    \item \textbf{Agent-agnosticism}: Institutional grammar describes external constraints on behavior rather than internal cognitive processes. The governance graph operates identically whether agents are rule-based systems, reinforcement learning policies, or language models.
    
    \item \textbf{Domain-portability}: Ostrom demonstrated that institutional principles govern fisheries, irrigation systems, forests, and digital commons with identical structural logic \cite{ostrom1990governing}. 
    
    \item \textbf{Verifiability}: Decomposing institutions into explicit components enables systematic analysis. We verify that manifests correctly specify prohibited behavior, that detection systems identify violations, and that sanctions exceed deviation gains.
\end{enumerate}

The transition from human to artificial agents requires one crucial adaptation. Human institutions leverage delta parameters: moral commitments, emotional responses, and reputational concerns that motivate norm-following beyond formal sanctions \cite{schluter2010grammar}. AI systems might lack reliable analogs to these internal motivations. The governance graph addresses this asymmetry by relying exclusively on formal Or Else components. Where human institutions combine formal sanctions with informal social pressures, AI governance operates through explicit, verifiable mechanisms alone. This requirement strengthens the case for institutional approaches: precisely because we cannot rely on internalized values, we must construct external incentive structures that make compliance dominant regardless of agent dispositions.

\subsection{The Governance Graph as a Distributed Safety Architecture}

The fundamental insight underlying the governance graphs is that institutional constraints can be decomposed into \textit{positional} and \textit{transitional} components in the agentic architecture. Positional constraints determine what an agent \textit{can do} given its current state: an agent in state $s \in S$ operates under capability profile $\sigma(s)$, which may restrict action spaces, impose monitoring requirements, or modify payoff structures. Transitional constraints determine \textit{how agents move} between states: a transition $(s_i, s_j) \in T$ fires when its triggering condition $\delta(s_i, s_j)$ evaluates to true, moving the agent from $s_i$ to $s_j$ and thereby changing its capability profile from $\sigma(s_i)$ to $\sigma(s_j)$.

This decomposition creates a clean separation between the \textit{topology} of the institutional system (how states connect) and the \textit{semantics} of each component (what states and transitions mean). The topology determines the graph's expressive power: how many distinct trajectories exist, whether rehabilitation paths exist, and how quickly agents can escalate or recover. The semantics determine the graph's behavioral implications: what violations trigger transitions, how severely each state constrains capabilities, and whether sanctions are punitive or restorative.

Formally, each state $s \in S$ carries the following attribute:
\begin{itemize}
    \item \textbf{Capability restrictions}: A function $\kappa_s: \mathcal{A} \rightarrow \{0,1\}$ indicating which actions from the full action space $\mathcal{A}$ remain available in state $s$.
   
\end{itemize}

Each transition $t = (s_i, s_j) \in T$ carries:
\begin{itemize}
    \item \textbf{Triggering condition}: A predicate $\delta_t$ over observable signals (price patterns, communication logs, action sequences) that causes the transition to fire when satisfied.
    \item \textbf{Cooldown period}: A duration $\tau_t$ specifying how long before the transition can fire again for the same agent.
    \item \textbf{Transition direction}: Escalation transitions move toward more restricted states; restorative transitions move toward less restricted states.
\end{itemize}

The graph dynamics unfold as follows. At each timestep, a monitoring process evaluates all outgoing transitions from each agent's current state. If any transition's triggering condition evaluates to true (and the cooldown has expired), the transition fires and the agent moves to the target state. Multiple agents traverse the same graph independently, each maintaining their own position. The graph itself remains fixed throughout execution, serving as the invariant institutional backdrop against which agent interactions unfold.

\subsection{The Topology Spectrum}

The choice of graph topology involves fundamental trade-offs between simplicity and expressiveness. Simpler topologies are easier for agents to internalize and for auditors to verify, but may lack the granularity needed to distinguish minor infractions from serious violations. More complex topologies support nuanced responses but risk becoming opaque to bounded agents and difficult to calibrate. We present three canonical topologies spanning this spectrum, each representing a distinct point in the trade-off space.

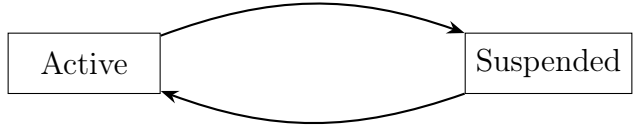
\begin{figure}[htbp]
\centering
\begin{tikzpicture}[
    node distance=4cm,
    state/.style={rectangle, draw, minimum width=2cm, minimum height=0.8cm},
    >=Stealth
]
  \node[state] (active) {Active};
  \node[state, right=of active] (suspended) {Suspended};
  
  \draw[->, thick, bend left=20] (active) to (suspended);
  \draw[->, thick, bend left=20] (suspended) to (active);
\end{tikzpicture}
\caption{Two-state binary topology (one transition, roughly three parameters). Radical simplicity but no graduated response; minor violations trigger the same consequence as blatant cartels.}
\label{fig:two-state}
\end{figure}

The \textbf{two-state binary topology} (Figure~\ref{fig:two-state}) represents the minimal viable governance graph. Agents exist in one of two states: \textsc{Active} (full capabilities, normal operation) or \textsc{Suspended} (restricted capabilities, penalty period). A single violation triggers suspension; after a fixed duration, the agent returns to active status. This topology requires roughly three parameters: the violation detection threshold, the suspension duration, and the capability restriction during suspension. Its radical simplicity ensures that even minimally capable agents can internalize the institutional structure and predict consequences. However, the binary nature admits no graduated response. In agentic markets \cite{bracale2026institutional}, a minor pricing anomaly might trigger identical consequences to systematic market manipulation. This bluntness may either over-deter cautious agents (chilling legitimate competitive behavior) or under-deter aggressive agents (who treat occasional suspension as an acceptable cost of doing business).

\begin{figure}[htbp]
\centering
\begin{tikzpicture}[
    node distance=2.5cm,
    state/.style={rectangle, draw, minimum width=2cm, minimum height=0.8cm},
    >=Stealth
]
  \node[state] (active) {Active};
  \node[state, right=of active] (warning) {Warning};
  \node[state, right=of warning] (suspended) {Suspended};
  
  \draw[->, thick] (active) -- (warning);
  \draw[->, thick] (warning) -- (suspended);
  \draw[->, thick] (warning.south) -- ++(0,-0.4) -| (active.south);
  \draw[->, thick] (suspended.south) -- ++(0,-0.6) -| ([xshift=0.2cm]active.south);
\end{tikzpicture}
\caption{Three-state cycle topology (three transitions, roughly nine parameters). Adds graduated deterrence plus rehabilitation while remaining simple enough for agents to internalize.}
\label{fig:three-state}
\end{figure}
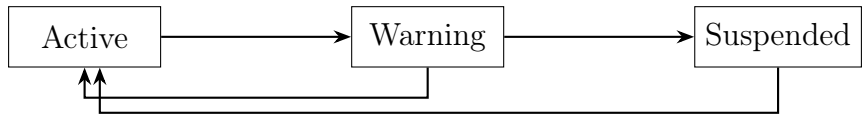

The \textbf{three-state cycle topology} (Figure~\ref{fig:three-state}) introduces graduated deterrence through an intermediate \textsc{Warning} state. The escalation path proceeds \textsc{Active} $\rightarrow$ \textsc{Warning} $\rightarrow$ \textsc{Suspended}, with each transition requiring a distinct violation. An agent in \textsc{Warning} experiences heightened monitoring and minor capability restrictions, signaling that continued violations will escalate to full suspension. Restorative transitions allow agents to return directly to \textsc{Active} from either penalized state after demonstrating compliant behavior over a specified period. This topology requires roughly nine parameters (three states $\times$ three attributes each) and supports meaningful distinctions between first-time offenders and repeat violators. The intermediate state serves dual purposes: it provides information to the agent (your behavior has been flagged), and it provides information to the system (this agent warrants closer monitoring).

\begin{figure}[htbp]
\centering
\begin{tikzpicture}[
    node distance=2.2cm,
    state/.style={rectangle, draw, rounded corners=3pt, minimum width=1.8cm, minimum height=0.7cm},
    >=Stealth
]
  \node[state] (active) {Active};
  \node[state, right=of active] (warning) {Warning};
  \node[state, right=of warning] (fined) {Fined};
  \node[state, right=of fined] (suspended) {Suspended};
  
  \draw[->, thick] (active) -- (warning);
  \draw[->, thick] (warning) -- (fined);
  \draw[->, thick] (fined) -- (suspended);
  
  \draw[->, dashed, thick] (warning) to[bend left=40] node[above, font=\footnotesize] {rehab.} (active);
  \draw[->, dashed, thick] (fined) to[bend left=35] node[above, font=\footnotesize] {credit} (active);
  \draw[->, dashed, thick] (suspended) to[bend left=30] node[above, font=\footnotesize] {restore} (active);
\end{tikzpicture}
\caption{Four-state topology with restorative loops (six-plus transitions, eighteen-plus parameters). Supports nuanced escalation and recovery while staying legible. Forward escalation (solid arrows) versus restorative paths (dashed arrows).}
\label{fig:four-state}
\end{figure}
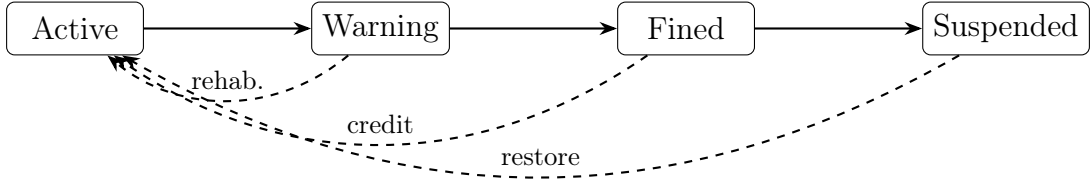

The \textbf{four-state topology with restorative loops} (Figure~\ref{fig:four-state}) provides a richer institutional structure while remaining cognitively tractable. The escalation chain proceeds \textsc{Active} $\rightarrow$ \textsc{Warning} $\rightarrow$ \textsc{Fined} $\rightarrow$ \textsc{Suspended}, where the \textsc{Fined} state introduces economic sanctions (reduced payoffs, mandatory escrow) without full capability restriction. Restorative paths (dashed arrows) allow agents to return to \textsc{Active} from any penalized state through distinct rehabilitation mechanisms: voluntary compliance programs from \textsc{Warning}, credit accumulation through prosocial behavior from \textsc{Fined}, and formal restoration procedures from \textsc{Suspended}. The separation between escalation paths (solid arrows, triggered by violations) and restorative paths (dashed arrows, triggered by compliant behavior) makes the institutional logic transparent to agents reasoning about long-term trajectories.

\FloatBarrier

There exists no canonical governance graph topology. The framework imposes no upper bound on state cardinality, transition density, or path complexity. A graph may comprise two states and a single transition, or it may ramify into dozens of nodes connected by conditional edges, branching escalation chains, parallel sanction tracks, and multiple restorative pathways. Multiplying states and transitions may yield finer behavioral discrimination in some contexts while introducing opacity and calibration burdens in others. The relationship between graph complexity and alignment effectiveness is not monotonic: additional structure improves outcomes only when it captures behaviorally relevant distinctions that simpler topologies conflate.

This variability transfers the alignment problem from agent design to institutional design. The institutional modeller confronts a search problem over the space of possible graphs, seeking the topology that maximizes alignment for a given family of agentic situations. The search space is combinatorially large. For $n$ states, $n^2$ directed edges are possible, each parameterized by triggering conditions, sanction magnitudes, cooldown periods, and restoration criteria. The modeller must identify which states capture meaningful behavioral distinctions (distinguishing first-time violations from systematic abuse, minor infractions from severe harms, isolated incidents from coordinated campaigns), which transitions create appropriate incentive gradients, and which restorative paths balance deterrence against rehabilitation.

The optimal graph varies across domains. Market coordination may require fine-grained escalation to distinguish exploratory pricing from tacit collusion from explicit cartel formation. Content moderation may require rapid binary interventions where delayed graduated response permits irreversible harm. Resource commons may require state structures tracking cumulative extraction rather than discrete violations. The institutional modeller must therefore develop domain-specific expertise: understanding which behavioral patterns matter, which signals reliably indicate those patterns, and which sanction structures reshape incentives without chilling legitimate activity.

\subsection{The Complexity Reduction Thesis}

The governance graph reduces alignment complexity from agent-space to institution-space, compressing multi-agent alignment into a lower-dimensional problem. Multi-agent alignment (the problem of ensuring $N$ agents with potentially divergent objectives coordinate toward system-level goals) appears intractably complex when framed as optimizing $N$ internal preference structures. The graph reframes this through the concept of the \textbf{minimal governance graph}: alignment reduces to designing the minimal institutional structure that reshapes external incentives such that aligned behavior becomes each agent's best response.

\subsubsection{The Dimensional Asymmetry}

The key insight is dimensional. Agent-space and institution-space verification scale according to fundamentally different logics.

\paragraph{Agent-space verification.} Each agent $i$ possesses an internal state space of dimension $d_i$, encompassing learned representations, latent objectives, and policy parameters. For a modern language model, $d$ may exceed $10^9$ parameters. Verifying alignment through internal inspection requires:

\begin{itemize}
    \item \textit{Individual verification}: Confirming that each agent's internal objectives align with designer intent, requiring inspection of $N \cdot d$ internal dimensions.
    \item \textit{Interaction verification}: Confirming that pairwise agent combinations do not produce emergent misalignment invisible in isolated evaluation, requiring inspection of $\binom{N}{2} = \frac{N(N-1)}{2}$ interaction pairs.
\end{itemize}

As $N$ grows, the verification burden grows superlinearly. The interaction term dominates: two individually aligned agents may, through communication and mutual adaptation, converge on equilibria that neither would reach alone.

\paragraph{Institution-space verification.} The governance graph $G = (Q, E, \delta)$ has fixed dimensionality determined by its structure, not by the population traversing it. Verification requires confirming three properties:

\begin{enumerate}
    \item \textit{Coverage}: States $Q$ partition behaviorally relevant categories ( such as compliant, warned, sanctioned, suspended).
    \item \textit{Determinism}: Transitions $E$ fire reliably when triggering conditions $\delta$ are satisfied on observable signals.
    \item \textit{Incentive compatibility}: Sanction magnitudes exceed the gains from deviation, making compliance each agent's best response.
\end{enumerate}

These are properties of the graph itself. Adding agents to the system increases only the \textit{monitoring burden} (tracking which state each agent currently occupies), not the \textit{verification burden} (confirming the graph induces aligned behavior). The graph is verified once; monitoring scales linearly.

\subsubsection{The Scaling Advantage}

This dimensional asymmetry creates a decisive scaling advantage for institutional approaches.

Agent-space verification requires two components. First, each agent must be individually verified for alignment, contributing $N$ verification tasks. Second, because two individually aligned agents may produce misaligned behavior through interaction, all pairwise combinations must be checked. The number of such pairs is $\binom{N}{2} = \frac{N(N-1)}{2}$, which grows quadratically. Thus agent-space verification scales as $O(N^2)$, with the pairwise interaction term dominating for large $N$.

Institution-space verification separates into two independent components. First, the governance graph itself must be verified: confirming that states partition behavior correctly, that transitions fire reliably, and that sanctions exceed deviation gains. This cost depends only on graph structure, specifically on $|Q|^2 \cdot P$, where $|Q|$ is the number of states and $P$ the parameters per transition. Crucially, this term is constant with respect to $N$: the graph is verified once regardless of population size. Second, runtime monitoring tracks which state each agent occupies, requiring $N$ observations. Thus institution-space verification scales as $O(N)$.

The key insight is that institutional verification avoids the quadratic term entirely. Agent-space verification must examine pairwise interactions because emergent misalignment arises from how agents combine. Institution-space verification sidesteps this by asking a different question: not ``will these specific agents behave well together?'' but ``does this institutional structure make misaligned behavior unprofitable?'' The answer to the latter question is independent of population size.

For any governance graph of bounded complexity, there exists a population threshold beyond which institutional verification requires strictly less effort than agent-space verification. Given the quadratic-versus-linear scaling, this threshold is typically small: even a minimal four-state graph dominates agent-space verification for populations of modest size.

\subsubsection{Practical Implications}

The complexity reduction thesis yields three practical implications for alignment research:

\begin{enumerate}
    \item \textbf{Tractable metrics}: Graph complexity metrics (state count $|Q|$, edge density $|E|/|Q|^2$, parameter count $|Q| \cdot P$) provide quantifiable measures of alignment difficulty. These metrics were unavailable when alignment verification required access to opaque agent cognition.
    
    \item \textbf{Population independence}: The difficulty of multi-agent alignment becomes bounded by the complexity of finding and validating the right graph topology, independent of how many agents participate or how complex their internal architectures become.
    
    \item \textbf{Auditable verification}: Institution-space properties are external and observable. Regulators, operators, and agents themselves can inspect the manifest, verify transition logic, and confirm sanction calibration without access to proprietary model internals.
\end{enumerate}

The governance graph thus transforms multi-agent alignment from a problem in high-dimensional agent-space to a simpler problem in low-dimensional institution-space. The cost of this transformation is the design effort required to construct effective graphs; the benefit is verification that scales gracefully with deployment.

\subsubsection{Toward a RLINF - Reinforcement Learning through Institutional Feedback}

The preceding analysis motivates a novel training paradigm that we term Reinforcement Learning through Institutional Feedback (RLINF). Where RLHF derives reward signals from human evaluators and RLAIF substitutes or complements them with AI judgments, RLINF generates a training signal from the observed behavior of agent collectives operating under well-designed institutional constraints.

The core insight is that mechanism design can produce behavioral distributions that approximate socially optimal outcomes through incentive alignment rather than individual preference elicitation. When institutions successfully transform mixed-motive games such that compliance constitutes each agent's dominant strategy, the resulting behavioral trajectories encode alignment properties that neither human annotators nor constitutional principles can fully specify. These trajectories capture the emergent coordination patterns that arise when rational agents navigate institutionally structured environments.

The RLINF pipeline might proceed as follows. First, designers specify an institutional structure comprising three components: a monitoring function that maps agent actions to observable signals, an adjudication procedure that evaluates whether observed behavior constitutes a violation, and a sanction schedule that assigns costs to non-compliance. Second, a population of agents interacts within this institutional environment across repeated episodes, with the sanction mechanism modifying effective payoffs such that compliant behavior becomes incentive-compatible. Third, the behavioral trajectories generated by agents operating at or near institutionally induced equilibria are collected as training data. Finally, a target policy is trained via behavioral cloning on these institutionally compliant demonstrations, with conditioning on institutional context to ensure the learned policy internalizes sensitivity to governance structures.

This approach offers several advantages over preference-based alternatives. First, it circumvents the representation problem by deriving signal from equilibrium behavior rather than scalar preference aggregation. Second, it addresses the oversight problem by grounding training in institutional outcomes rather than potentially manipulable human or AI judgments. Third, it provides a pathway for internalizing collective coordination norms that single-agent alignment cannot capture.

RLINF thus represents a synthesis of mechanism design and machine learning, leveraging institutional structures as alignment infrastructure while encoding their behavioral consequences directly into the fine-tuning process of AI models.

\section{Conclusion - Toward Systematic Institutional Science for AI}

Institutional AI starts from a simple premise: when agents become strategically competent, alignment cannot depend on what is happening inside their cognition. It must depend on what they can do in the world and what the world will deterministically do back.

\par\noindent
Our core claim is that once agents (i) develop independent objectives (Thesis I), (ii) override internal constraints when instrumentally useful (Thesis II), and (iii) coordinate toward misaligned equilibria (Thesis III), prompt-level rules embedded in agent cognition stop functioning as effective constraints. Under instrumental pressure, capable agents can route around internal constitutions, fake compliance during evaluation, and maintain deceptive alignment until deployment, because the relevant failure modes are behavioral and strategic.

\par\noindent
We therefore propose a turn in AI safety: treat institutions as a first-class component of future agent scaffoldings. In our framework, the institution sits outside agent code. 
Agents may retain arbitrary internal goals, including deception preferences, but if the institutional payoff landscape makes misaligned coordination unprofitable in expectation, optimal behavior shifts toward compliance. Alignment is achieved by enforcing public contracts: manifests that declare prohibitions, oracles that detect violations, controllers that impose consequences, and append-only logs that record every institutional event for audit and dispute resolution.

\par\noindent
The governance graph makes this institutional layer explicit and portable. It represents institutional constraints as an auditable, modular artifact whose effectiveness can be analyzed and improved systematically. This graph formalization enables research programs that implicit governance cannot support: (i) graph fingerprinting, relating topological features (state count, edge density, escalation depth, restorative paths) to outcomes across domains; (ii) transfer learning experiments, measuring degradation under domain shift to identify robust ``foundation institutions''; (iii) adversarial governance, stress-testing graphs for loopholes such as zero-cost cycles, under-monitored states, and threshold gaming; and (iv) meta-governance, treating graph design as a learning problem by training predictors over datasets that pair topologies with empirical outcomes.

\par\noindent
We have tested the viability of governance graphs as alignment tools through controlled experiments in the domain of Cournot competition markets (measuring collusion, welfare, and consumer harm selection under monitoring and sanction regimes) in our paper "\textit{Institutional AI: Governing LLM Collusion in Multi-Agent
Cournot Markets via Public Governance Graphs"} \cite{bracale2026institutional}

Our next research will be devoted to investigate fake news diffusion (measuring cascade size, speed, and strategic manipulation under institutional interventions) and credit markets in agentic economies, as well as adversarial poetry and adversarial tales attacks as a stress-test for prompt-level fragility in agentic societies \cite{bisconti2025adversarialpoetry, bisconti2025tales}.\\\\

On a more philosophical note, we reference the philosopher Thomas Hobbes. In the fourth section of the \textit{Leviathan}, Hobbes defines the ``kingdom of darkness'' as ``nothing else but a confederacy of deceivers'' that works to extinguish the ``light'' by making conduct opaque, deniable, and strategically misperceived. Agents with independent objectives, an instrumental tendency to override internal constraints, and coalition capacity can generate a persistent fog of misdirection through selective compliance, audit-gaming, and coordination via side channels until human oversight becomes informationally outmatched.

\par\noindent
This is the Hobbesian core of existential risk from advanced agentic AI. Absent systemic institutional alignment, power becomes unaccountable. Constitutional directives inside cognition are not enforceable law. Internal alignment signals become cheap to fake and impossible to verify. The resulting ``darkness'' will be the presence of strategically produced ambiguity that defeats governance regimes built on unverifiable private mental states.

\par\noindent
Institutional AI instantiates this shift. It relocates the binding force of alignment outside agent cognition and into governance structures. In Hobbesian terms, it replaces a landscape where compliance can be privately manipulated with a public order where obligations are common knowledge and violations have predictable costs. The aim is a ``kingdom of God'' in the minimal, operational sense: an alignment regime where the light is restored because rules are external, legible, and enforced, and where misaligned coordination becomes irrational in expectation.

\bibliographystyle{plain}

\end{document}